**Observing Colossal and Tunable Near-field Thermal Radiation with A Highly Sensitive Annular Micro-thermocouple Junction**

*Ken Araki,[#] Wei Han Won,[#] and Liping Wang**

K. Araki, L. Wang
Mechanical and Aerospace Engineering, Fulton Schools of Engineering, Arizona State University, Tempe, Arizona 85287, USA

W. H. Won, L. Wang
Materials Science and Engineering, Fulton Schools of Engineering, Arizona State University, Tempe, Arizona 85287, USA

[#] Equal contributions.
[*] Corresponding author. E-mail: liping.wang@asu.edu

Funding: U.S. National Science Foundation (CBET-2212342 and ECCS-2309663) and U.S. Department of Energy (DE-SC0024201)

**Abstract**: Near-field radiative heat transfer between two objects has been theoretically predicted and experimentally demonstrated to exceed far-field blackbody limit across nanoscale vacuum gap distance enabled by evanescent surface waves coupling, while significant enhancement by more than 100 times usually requires sub-50-nm vacuum gaps around room temperature. This is challenging for parallel-plate configuration with millimeter sample sizes due to intrinsic wafer bow and contaminant particles. Sphere-plate configuration with a microsphere attached to bimaterial cantilevers or micro-thermocouple tips has been used to experimentally demonstrate near-field radiative heat transfer down to 30-nm gaps, but it is much less developed because of the challenges in the sophisticated sensor fabrication, low sensitivity and weak signals. In this work, we overcome these challenges by an annular micro-thermocouple junction fabricated at the end of a glass fiber with straightforward thin-film deposition to achieve high measurement accuracy with large Seebeck coefficient 25 μV/K and thermal resistance $8.7\times10^6$ K/W. With a silica microsphere attached underneath the micro-thermocouple junction, we report experimental observation of colossal near-field radiation heat transfer over blackbody limit down to 10-nm gap up to 2600 times with quartz and 900 times with doped silicon. Upon phase transition of $VO_2$ thin film emitter, tunable near-field heat transfer up to 430-fold enhancement is experimentally demonstrated at 15-nm gap with 64% reduction. Experimental data agrees well with rigorous modeling based on fluctuational electrodynamics and Derjaguin approximation, and underlying mechanism is understood by energy transmission calculations. The results will advance the experimental study and fundamental understanding of energy transport at nanoscale gaps.

## 1. Introduction

As near-field radiative heat transfer could surpass the far-field blackbody limit by several orders of magnitudes at nanometric vacuum gaps, it has spurred the immense theoretical studies in nanoscale energy transport for a variety of applications such as solid-state energy conversion, photoluminescent refrigeration, and near-field imaging and manufacturing.[1–5] Experimentalists have successfully developed different approaches to measure the near-field thermal radiation.[6] Plate-plate configuration has enabled the experimental studies of different materials at the sub-micron vacuum gaps either created by spacers[7–14] or maintained by precise motion control[15–20]. However, it is hard to achieve sub-100-nm gaps with millimeter-sized samples or sub-50-nm with microscale samples due to intrinsic wafer bow and contaminant particles, and over 100-fold enhancements beyond blackbody limit was barely reported.[21] With costly fabrication of custom thermal sensors, where nanoscale thermocouple junction is formed at the probe apex, tip-surface radiative heat transfer and energy transport in the extreme near field has been experimentally investigated across several nm to sub-nm gaps,[22–27] but it restricts the measurements at larger gaps due to weak signals and limits broad selection of materials.

By attaching a microsphere onto microscale thermal sensors such as bimaterial cantilevers, resistive thermometers, and thermocouple junctions, sphere-plate configuration is used in several pioneering works that successfully observed strong near-field enhancement down to 30-nm gaps limited by snap-in or bending in closer proximity[28–36]. Giant near-field radiation enhancement in the sub-30-nm gap regime is rarely reported from direct experiments[37]. Sphere-plate near-field measurement has been much less developed over last decade compared to the widely used plate-plate counterpart. After all, the reported microsphere temperature varied less than 1 K or even 0.1 K during the near-field measurements with bimaterial cantilevers or micro-thermocouples due to small thermal resistance and low sensitivity, which poses a great challenge in detecting low signals in addition to sophisticated sensor fabrication and precise nanometric gap control.

Tunable near-field radiative heat transfer has been theoretically studied for designs of near-field thermal rectifier (or diode)[38–41], thermal switch, thermal modulator/transistor, and thermal memory with thermochromic $VO_2$[42–49], electrically-gated graphene[50–52], magnetic field tuned magneto-optical materials[53–56], mechanically-deformed thin films[46,57] or twisted nanostructures[55,58]. To date there have been several experimental demonstrations of tunable near-field radiation with graphene down to 55-nm gap[59–61] and with $VO_2$ down to 60-nm gaps[32,62,63], but that at sub-50-nm vacuum gaps with colossal near-field enhancement has not been experimentally reported.

This work aims to demonstrate the observation of colossal and tunable near-field radiative heat transfer with a silica microsphere attached underneath a glass fiber end where a highly sensitive annular micro-thermocouple (μ-TC) junction is fabricated. The Seebeck coefficient of fabricated μ-TC samples is carefully calibrated, and its thermal resistance is fitted with the far-field data from rigorous thermal modeling. Quartz and heavily doped silicon (HDSi) emitters are measured at a wide range of vacuum gaps down to 10 nm for high thermal conductance and near-field heat transfer coefficient with orders of magnitude exceeding blackbody limit. Thermochromic $VO_2$ thin film is tested to unveil the tunable radiative heat transfer at tens of nanometer gaps with different phases. Radiative heat transfer based on fluctuational electrodynamics and Derjaguin approximation is calculated to validate the near-field measurements, and photon tunneling probability is presented to elucidate the mechanism.

## 2. Results and Discussion

### 2.1. Design and Fabrication of the Annular Micro-Thermocouple (μ-TC) Junction

As depicted in Figure 1A, the μ-TC is fabricated on a thin glass fiber of 75 μm in diameter with a length of 25 mm in order to achieve high thermal resistance. One end as the cold junction with temperature fixed at $T_0$ = 25°C is glued onto a copper-coated microscope cover slip for easy handling and wire connection, and an annular thermocouple junction is fabricated on the other free end where a 100-μm-diameter silica microsphere is attached. With some careful preparations for the coreless glass fiber and the cover slip (Figure S1A for details), 100-nm chromium (Cr), 8-μm Parylene C, and 100-nm nickel (Ni) are sequentially coated onto the glass fiber and cover slip as illustrated in Figure 1B.

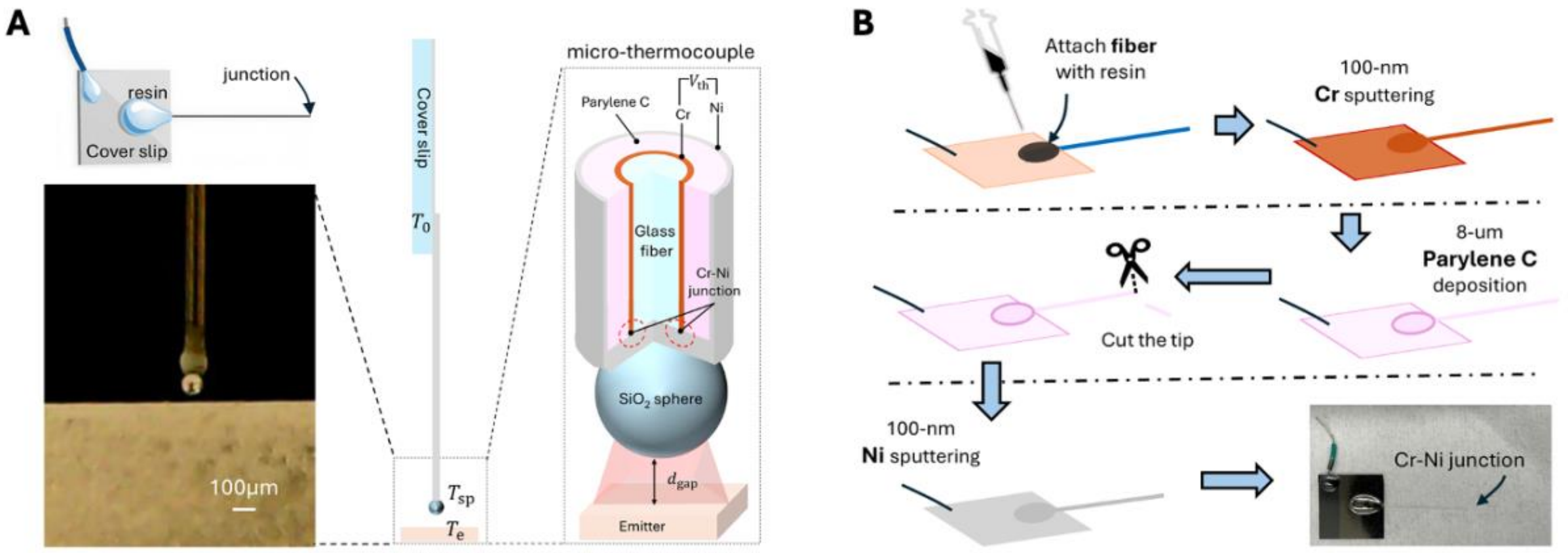

**Figure 1. A**. Design and photo of annular micro-thermocouple (μ-TC) junction formed at the end of a thin glass fiber with a silica microsphere attached for near-field thermal radiation measurements. **B**. Fabrication process for the annular μ-TC junction involving thin film depositions of chrome (Cr), Parylene C, and nickel (Ni).

The Parylene C layer electrically insulates the two metallic layers everywhere except at the free end of the glass fiber. This is realized by delicately cutting the fiber tip after the deposition of Parylene C to expose the annular shape of Cr only at the very end surface of the glass fiber, which forms the μ-TC junction after the sputtering of Ni, whereas Cr and Ni are selected for the thermocouple materials for the purpose of large Seebeck coefficient. Different from other works where the microspheres were attached to the side of bimaterial cantilever tips[28–34] or μ-TC tip[36], the silica microsphere is glued onto the μ-TC junction right underneath the glass fiber end (Figure S1B for details), possibly allowing to reach sub-30-nm gaps without bending or snap-in. Considering large annular junction area instead of a sharp tip, uniform microsphere temperature ($T_{sp}$) is reasonably assumed and directly measured by the μ-TC junction. During the near-field thermal radiation measurement, an emitter sample heated at temperature $T_e$ is placed underneath the silica microsphere in high vacuum with gap distance ($d_{gap}$) precisely controlled with a closed-loop piezoelectric nanopositioner at nm resolutions from tens of micrometers down to 10 nm.

## 2.2. Calibration of Seebeck Coefficient and Fitting for Thermal Resistance

By immersing the bare fiber end without microsphere into a pool of thermal paste on a heater stage, the Seebeck coefficients of fabricated μ-TC samples of different metal layers are carefully calibrated by measuring the μ-TC voltages at different temperatures (Figure S2 for details). As shown in Figure 2A, the μ-TC samples with junction formed by Cu and Ni, it yields a Seebeck coefficient of 11 μV/K. With Cr – Ni junction, the Seebeck coefficient increases to 20 μV/K. With a pre-deposited copper layer on the cover slip to enhance the electrical connection of Cr, the final Cu/Cr – Ni junction reaches a Seebeck coefficient of 25 μV/K, which is 5 times higher than the Pt – Au junction reported for similar sphere-plate near-field measurements, or 1.5~3 times sensitive than the nano-junction used for tip-plate near-field tests, as listed in Table 1.

**Table 1. Comparison of micro/nano-TC junctions for near-field radiation measurements**

| NFR format | TC-junction | Seebeck coefficient (μV/K) | Thermal resistance (K/W) | Ref. |
|---|---|---|---|---|
| Tip-plate | Pt - Au | 8 | $5.4 \times 10^4$ | Kittel, 2005 [22] |
| | Pt - Au | 9.5 | $1 \times 10^6$ | Wischnath, 2008 [23] |
| | Cr - Au | 16.28 | $1.6 \times 10^6$ | Kim, 2015 [24] |
| Sphere-plate | Pt - Au | 4.66 | $1.5 \times 10^6$ | Dang, 2025 [36] |
| | Cr - Ni | 25 | $8.7 \times 10^6$ | This work |

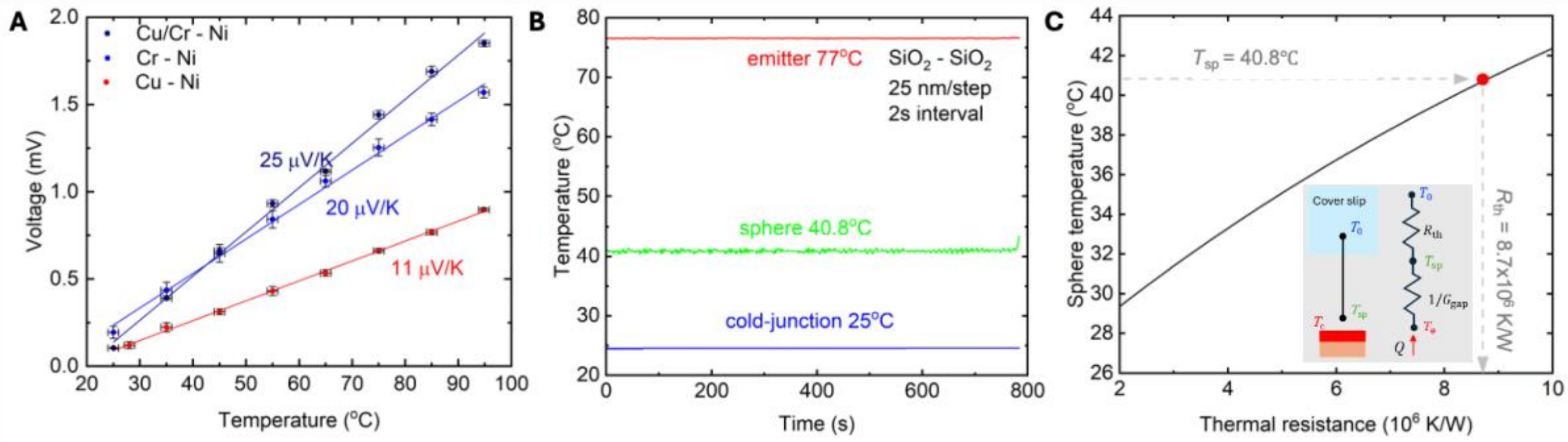

**Figure 2. A.** Calibrated Seebeck coefficients of fabricated micro-thermocouple (μ-TC) samples with different junction metal combinations including Cu/Cr – Ni, Cr – Ni, and Cu – Ni. Note that Cu layer is pre-deposited on the cover slip only to enhance the electrical contact of Cr for reliable voltage reading. **B**. Temperature profiles of the quartz emitter ($T_e$), silica microsphere ($T_{sp}$) attached to the μ-TC junction, and cold junction ($T_0$) during the initial approaching from far field about 10 μm away at 25 nm/step by a nanopositioner before reaching near field. **C**. Fitting of the μ-TC thermal resistance based on the microsphere temperature in the far field by comparing the measured value to theoretical modeling. Inset depicts the 1D thermal resistance network for the modeling.

To characterize the near-field radiative heat transfer, thermal resistance ($R_{th}$) of the coated glass fiber must be determined as well. Under high vacuum (<0.005 Pa) in the developed near-field setup (Figure S3 for details), the silica microsphere attached underneath the μ-TC junction is brought closer to a heated quartz wafer maintained at $T_e$=77°C by the nanopositioner from about 10 μm away to ~1 μm gap at a speed of 25 nm per 2 sec before entering the near-field regime ($d_{gap}$<1 μm). As shown in Figure 2B, the microsphere temperature $T_{sp}$ measured by the μ-TC junction maintains a constant value of 40.8°C in this far-field regime, where near-field effect is negligible. From the fluctuational electrodynamics and Derjaguin approximation, the rate of radiative heat transfer in the far field ($Q_{FF}$) between the silica microsphere and quartz wafer is calculated by considering only the propagating waves as a function of microsphere temperature. According to the 1D heat transfer model, the thermal resistance of the μ-TC is then correlated to the microsphere temperature by $R_{th} = (T_{\text{sp,FF}} - T_0)/Q_{FF}$ as shown in Figure 2C. With measured microsphere temperature of 40.8°C in the far field, the thermal resistance of the μ-TC is found to be $R_{th}$ = 8.7×10$^6$ K/W, which is 6 times larger than the micro-TC reported for the similar sphere-plate near-field tests, or 6 ~ 160 times greater than the nano-TC used for tip-plate near-field measurements. The fitted $R_{th}$ value is consistent with the theoretical estimate of that of bare glass fiber (~6×10$^6$ K/W), while the difference can be understood by the additional thermal resistance from the fixed fiber end to the thermistor for $T_0$ measurement. Larger thermal resistance could lead to greater microsphere temperature changes during the near-field radiation measurements for better signal/noise ratio and measurement accuracy.

### 2.3. Observation of Strong Near-field Radiation with Phononic and Plasmonic Emitters

With calibrated Seebeck coefficient and the fitted thermal resistance of the μ-TC, experimental thermal conductance can be found by $G_{gap} = [(T_{sp} - T_0)/(T_e - T_{sp})]/R_{th}$ based on measured microsphere temperature at different vacuum gap distances. Figure 3A presents the measured thermal conductance for the quartz emitter from 1 μm down to 10 nm gaps with multiple repeated scans, where the averaged microsphere temperature as a function of time could change by up to 9°C, which is one or two orders of magnitude higher than that in other reported sphere-plate near-field radiation tests[29,36]. The measured thermal conductance $G_{gap}$ shows excellent agreement with the modeling (Figure S4 and S5 for details) with large enhancement from ~52 nW/K at 1-μm gap to ~118 nW/K at 10-nm gap. In the similar way, the *p*-type HDSi emitter is also measured, where relatively smaller enhancement on the thermal conductance is observed from ~30 nW/K at 1-μm gap to ~50 nW/K at 10 nm, clearly validated by the modeling as shown in Figure 3B. Note that the near-field measurements show consistent results with different scan speeds and more repeated scans in Figure S6.

However, the thermal conductance includes the radiative heat transfer from both far-field and near-field contributions. As the far-field radiation does not change with vacuum gap distance in the near-field regime, the thermal conductance can be simply subtracted by its value at ~10-μm gap, and an equivalent sphere-plate near-field heat transfer coefficient ($h_{nfr}$) is obtained after the normalization with an equivalent area of $2\pi R d_{gap}$, where $R$ is the radius of the microsphere. Figure 3C and 3D present the measured $h_{nfr}$ in excellent agreement with the modeling for the quartz and *p*HDSi emitters, respectively, where strong near-field enhancement with sudden increase can be observed in the sub-100-nm gap regime. In particular with the quartz emitter, $h_{nfr}$ at 30-nm gap is measured to be 3.4±0.1 kW m$^{-2}$ K$^{-1}$, which matches the theoretical value of 2.6 kW m$^{-2}$ K$^{-1}$ in consistence with reported values at smallest gaps by others[29]. At 20-nm gap, the measured $h_{nfr}$ increases to 6.5±1.2 kW m$^{-2}$ K$^{-1}$ and up to 20.6±0.5 kW m$^{-2}$ K$^{-1}$ at 10-nm gap. With logarithmic scale, measured $h_{nfr}$ between the silica microsphere and the quartz reaches the blackbody limit ($h_{bb}$=0.008 kW m$^{-2}$ K$^{-1}$) at ~1μm gap and exceeds it at 10-nm gap by ~2600 times, demonstrating colossal enhancement experimentally observed for the first time for sphere-plate near-field radiation measurements, thanks to the careful design and high sensitivity of the fabricated μ-TC. While relatively smaller $h_{nfr}$ is experimentally observed for the *p*HDSi emitter, it still exceeds the blackbody limit by ~900 times with 7.2±1.5 kW m$^{-2}$ K$^{-1}$ measured at 10-nm gap (~800 times with 6.5 kW m$^{-2}$ K$^{-1}$ from modeling). The gap-dependent $h_{nfr}$ is nearly the same regardless the microsphere sizes (Figure S7), and the sphere-plate $h_{nfr}$ approximates well its plate-plate counterpart (Figure S8) from additional modeling.

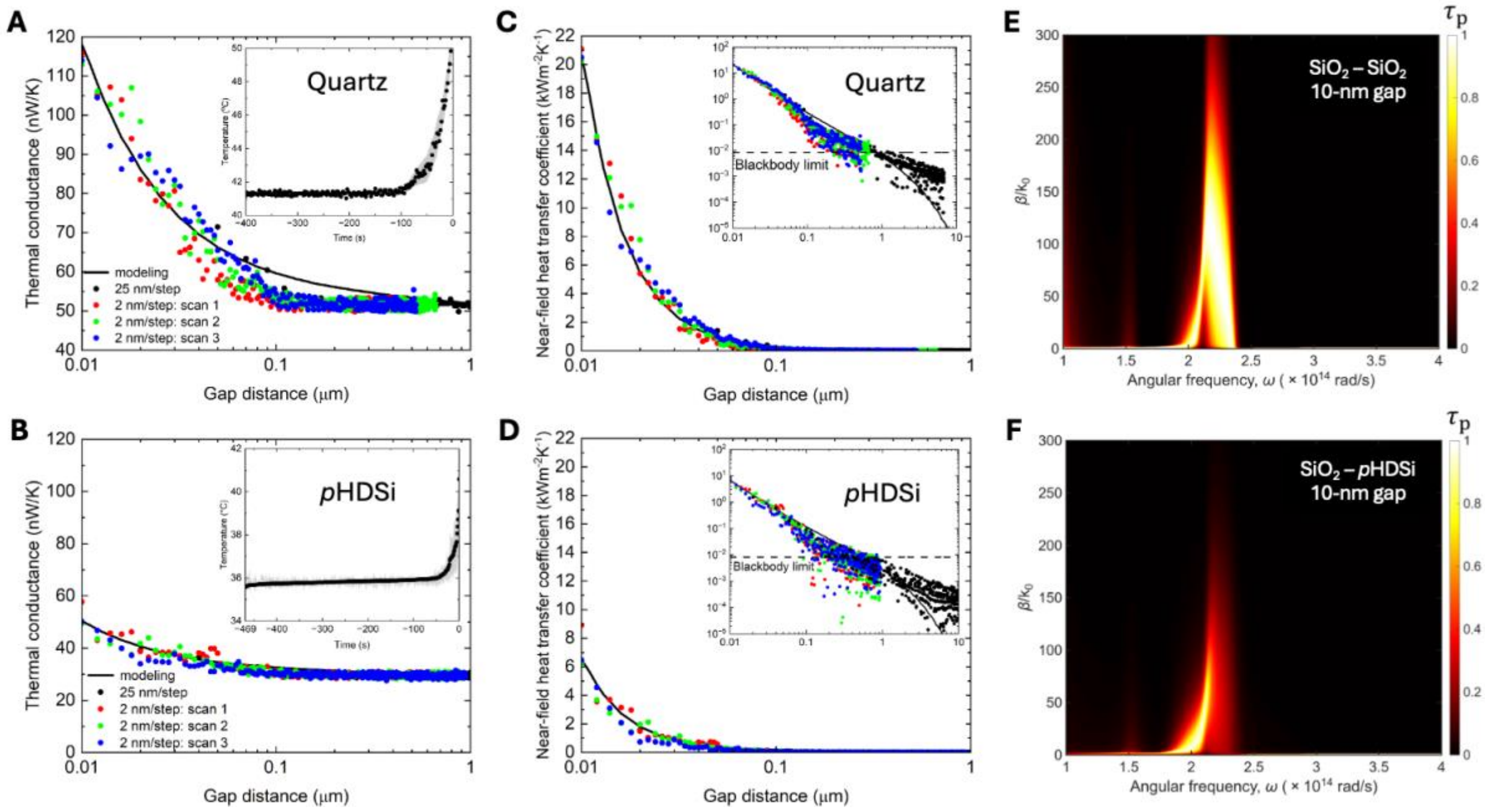

**Figure 3. A.B.** Measured overall thermal conductance ($G$) from multiple scans as a function of gap distance ($d_{gap}$) from 1 μm to 10 nm in comparison with the modeling with contributions from both far-field and near-field thermal radiation for (**A**) the quartz sample and (**B**) the p-type heavily-doped silicon (*p*HDSi) sample both maintained at 350 K. Insets show the microsphere temperature during approaching from far field to near field. **C.D.** Near-field heat transfer coefficient ($h_{nfr}$) in relation to the gap distance ($d_{gap}$) from both measurements and modeling for (**C**) quartz and (**D**) *p*HDSi. Insets present $h_{nfr}$ in the logarithmic scale in the larger vacuum gap distance with reference to the blackbody limit. **E.F.** Near-field energy transmission coefficient for *p*-polarized waves ($\tau_p$) as functions of angular frequency ($\omega$) and normalized parallel wavevector ($\beta/k_0$) from fluctuational electrodynamics calculation for the plate-plate configuration at 10-nm gap distance for (**E**) quartz and (**F**) *p*HDSi.

The strong near-field radiation enhancement is elucidated with the contour plots of energy transmission coefficient (or photon tunneling probability) for *p*-polarized waves only at 10-nm vacuum gap calculated from the fluctuational electrodynamics in the parallel-plate configuration for the quartz and *p*HDSi emitters in Figure 3E and 3F, respectively. Strong resonant coupling of phonon-phonon polaritons is clearly seen for $SiO_2$-$SiO_2$ system with greatly enhanced energy transmission with high-*k* modes (or evanescent energy channels) within the Reststrahlen band of $SiO_2$. The energy transmission becomes less both in frequencies and wavevectors when the $SiO_2$ receiver is paired with the *p*HDSi emitter due to weak resonant coupling of optical phonons in $SiO_2$ and plasmons in *p*HDSi. Please see Figure S9 for the optical constants used for radiative heat transfer and energy transmission calculations with verification from measured infrared properties for the quartz and *p*HDSi wafer samples.

### 2.4. Observation of Tunable Near-field Radiation with Thermochromic Emitter

A thermochromic emitter of 300-nm $VO_2$ thermally grown on the undoped silicon wafer is also measured for the near-field radiative heat transfer with the silica microsphere attached μ-TC. $VO_2$ is known to experience insulator-to-metal phase transition around 68°C, which has been widely used to develop tunable coatings for self-adaptive thermal control[64–66] or energy harvesting[67,68] with far-field thermal radiation. The temperature-depdendent infrared spectrometric characterization reveals excellent phase transition behaviors of this pristine $VO_2$ thin film grown via rapid thermal processing (RTP)[69] in Figure S10 along with fitted optical constatns for its insulating and metallic phases, which are used for near-field modeling. Figure 4A and 4B present the measured thermal conductance of the $VO_2$ thin film emitter maintained at 325 K in its insulating phase and at 350 K in its metallic phase, respectively. Evidently the measurement data from multiple scans match well with the modeling for both phases in the near-field regime with gap distances from 1 μm down to 15 nm. While the thermal conductance with insulating $VO_2$ is higher overall than that with metallic phase, which is due to larger far-field radiative heat transfer from higher emissivity in insulating phase, it increases in both cases at smaller vacuum gaps due to the stronger near-field effect. The thermal conductance is also measured at a fixed gap distance (such as 25 nm in Figure S12) with variable $VO_2$ temperatures, where sudden change upon phase transition along with thermal hystesis behavior between heating and cooling processes is clearly observed in the near field.

Figure 4C and 4D show the near-field heat transfer coefficient ($h_{nfr}$) respectively for the insulating and metallic $VO_2$, where excellent agreement is seen between the multiple measurements and the modeling. In particular, with the insulating $VO_2$ at 325 K, $h_{nfr}$ is experimentally measured to reach a value of 0.7±0.2 kW $m^{-2}$ $K^{-1}$ at 30-nm gap and 3.0±0.5 kW $m^{-2}$ $K^{-1}$ at 15-nm gap, exceeding the blackbody limit ($h_{bb}$ = 0.007 kW $m^{-2}$ $K^{-1}$) by ~130 and ~430 times, respectively. On the other hand, $h_{nfr}$ drops to 1.1±0.2 kW $m^{-2}$ $K^{-1}$ at 15-nm gap by 64% with metallic $VO_2$ at 350 K, where strong tunable near-field radiative heat transfer far beyond blackbody limit is experimentally observed upon phase transition. Calculation suggests at 10-nm gap that, higher near-field heat transfer coefficient $h_{nfr}$ of 6.3 kW $m^{-2}$ $K^{-1}$ (900-fold $h_{bb}$) with insulating $VO_2$ and 2.1 kW $m^{-2}$ $K^{-1}$ (300-fold $h_{bb}$) with metallic $VO_2$ can be achieved along with 68% reduction upon phase transition. The direct experimental observation of colossal and tunable near-field radiative heat transfer here undoubtedly confirms the great promise of $VO_2$ in modulating heat flow through nanometric vacuum gaps beyond blackbody limits by hundreds of folds in future thermophototronic and quantum applications.

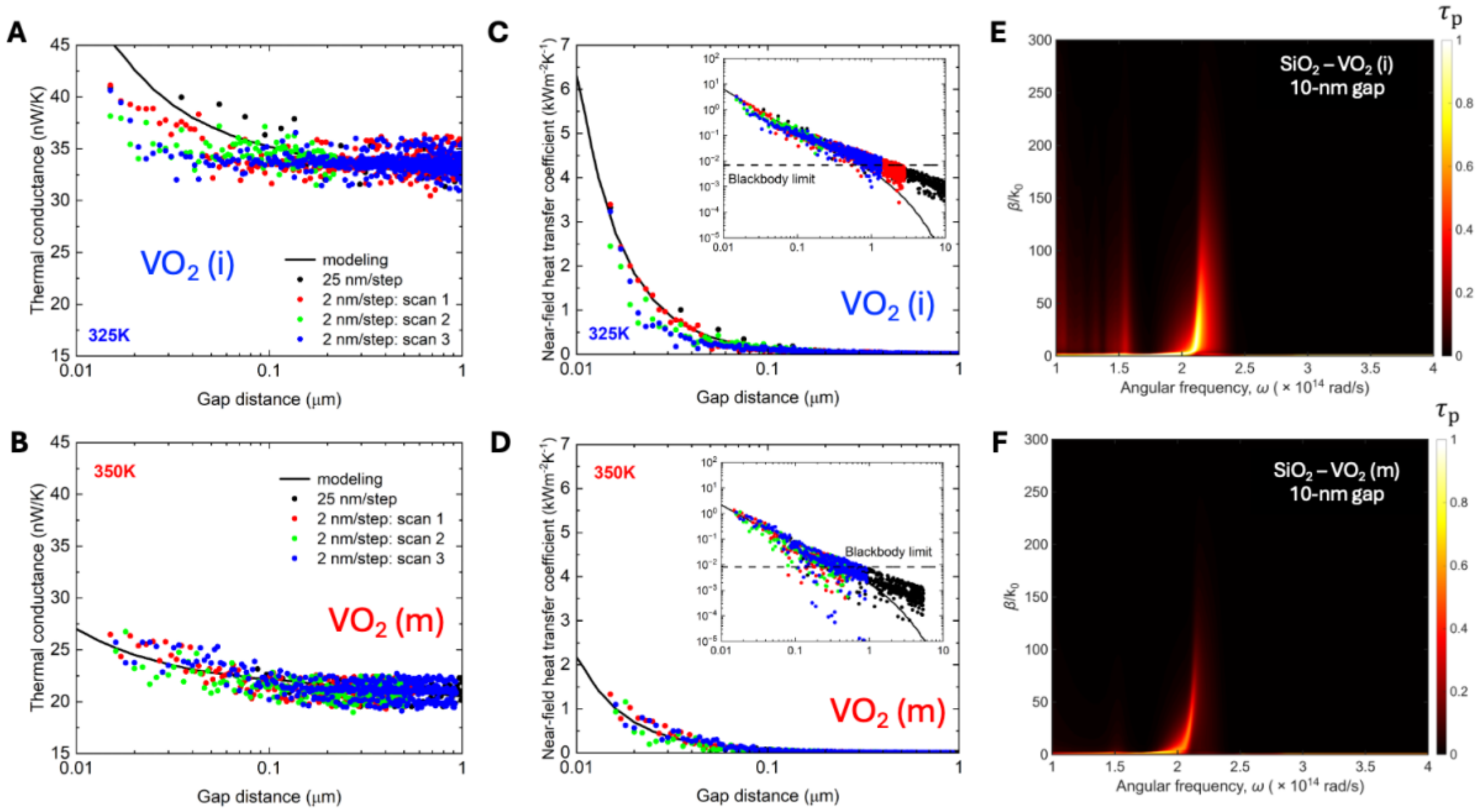

**Figure 4. A.B.** Measured overall thermal conductance ($G$) from multiple scans as a function of gap distance ($d_{gap}$) from 1 μm to 15 nm in comparison with the modeling with contributions from both far-field and near-field thermal radiation for the $VO_2$ thin film in (**A**) the insulating phase at 325 K and (**B**) the metallic phase at 350 K. **C.D.** Near-field heat transfer coefficient ($h_{nfr}$) in relation to the gap distance ($d_{gap}$) from both measurements and modeling for (**C**) the insulating and (**D**) metallic $VO_2$. Insets present $h_{nfr}$ in the logarithmic scale with reference to the blackbody limit. **E.F.** Near-field energy transmission coefficient from fluctuational electrodynamics calculation in the plate-plate configuration at 10-nm gap distance for the silica paired with the $VO_2$ thin film in (**E**) the insulating and (**F**) metallic phases.

The experimentally observed strong and tunable near-field radiative heat transfer between the $VO_2$ thin film emitter and the silica microsphere can be understood by the energy transmission calculated at 10-nm gap in the parallel-plate case for the insulating and metallic $VO_2$ as shown in Figure 4E and 4F, respectively. As the insulating $VO_2$ owns several phonon absorption modes in the infrared (see Figure S10B), which overlaps with those of $SiO_2$, resonant phonon-phonon coupling with large high-k modes occurs around the angular frequencies of $2.1\times10^{14}$ rad/s and $1.5\times10^{14}$ rad/s, where strongly enhanced energy transmission can be clearly seen. On the other hand, as $VO_2$ turns metallic, there is no strong resonant coupling with $SiO_2$ but non-resonant evanescent modes of $SiO_2$ phonons that channel photon energy across the nanometric gaps weakly. Note that this physical mechanism has been predicted by Yang et al. [42] with $SiO_2$ paired single-crystalline uniaxial $VO_2$ for strongly tunable near-field radiative thermal rectification but it has not experimentally observed due to challenges in both high-quality growth of $VO_2$ and near-field radiation measurements down to sub-50-nm until this work, owing to the RTP-grown pristine $VO_2$ and high sensitivity of fabricated μ-TC.

## 3. Conclusion

In summary, with a silica microsphere attached underneath a highly sensitive annular micro-thermocouple junction at the end of a glass fiber, we have experimentally demonstrated the colossal near-field radiative heat transfer down to 10-nm vacuum gap with up to 2600 times enhancement over blackbody limit from the quartz emitter and 900 times enhancement from the HDSi emitter. Tunable super-Planckian energy transport up to 430-fold enhancement with 64% drop at 15-nm gap is also directly observed upon phase transition of the $VO_2$ thin film emitter. The results will advance the experimental study of energy transport at nanoscale gaps with different material systems for understanding fundamental physics and enabling future engineering applications.

**Supporting Information**

Supporting figures S1 – S12 are included.

**Acknowledgements**

This work was supported by the U.S. National Science Foundation under Grant Nos. CBET-2212342 and ECCS-2309663, as well as and U.S. Department of Energy under Award No. DE-SC0024201. We would like to thank ASU NanoFab for use of their nanofabrication and characterization facilities.

**Conflict of Interests**

The authors have no conflict of interest to declare.

**Data Availability**

The data to support the findings of this work is available upon reasonable request.

**Author Contributions**

K.A. led the effort in developing the near-field test setup, while W.H.W. took care of most modeling work. Both K.A. and W.H.W. fabricated and calibrated μ-TC samples, conducted the near-field thermal radiation measurements, analyzed the data, prepared the figures, and drafted the initial manuscript. L.W. conceived the idea, supervised the work, secured the funding, and revised the manuscript. All authors have reviewed and approved the final version of the manuscript.